\documentclass[twocolumn,10pt,amsmath,amssymb,superscriptaddress,pra,aps]{revtex4-1}
\usepackage{graphicx}
\usepackage{bm}
\usepackage[T1]{fontenc}
\usepackage{hyperref}
\hypersetup{
 colorlinks  = true, 
 urlcolor   = black, 
 linkcolor  = blue, 
 citecolor  = red  
} 

\begin{document} 
\title{Signatures of unconventional pairing \\ in spin-imbalanced one-dimensional few-fermion systems}
\author{Daniel \surname{P{\k e}cak}}
\affiliation{\mbox{Faculty of Physics, Warsaw University of Technology, Ulica Koszykowa 75, PL-00662 Warsaw, Poland}} 
\author{Tomasz \surname{Sowi\'{n}ski}}
\affiliation{\mbox{Institute of Physics, Polish Academy of Sciences, Aleja Lotnikow 32/46, PL-02668 Warsaw, Poland}}
\date{\today}
\begin{abstract}
A system of a few attractively interacting fermionic $^6$Li atoms in one-dimensional harmonic confinement is investigated. Non-trivial inter-particle correlations induced by interactions in a particle-imbalanced system are studied in the framework of the noise correlation. In this way, it is shown that evident signatures of strongly correlated fermionic pairs in the Fulde-Ferrell-Larkin-Ovchinnikov state are present in the system and they can be detected by measurements directly accessible within state-of-the-art techniques. The results convincingly show that the exotic pairing mechanism is a very universal phenomenon and can be captured in systems being essentially non-uniform and far from the many-body limit.
\end{abstract}
\maketitle

\section{Introduction}
One of the cornerstones of our understanding of strongly correlated states of quantum matter is based on the theory of superconductivity by Bardeen, Cooper, and Schrieffer \cite{bcs1957}. In this theory, the existence of the superconducting phase is explained following the fundamental observation by Cooper \cite{Cooper1956} that the ground-state energy of an attractively interacting system is significantly decreased by the collective formation of Cooper pairs --- non-trivially correlated states of two fermions with exactly opposite momenta.
Based on this idea of collective pairing, a plethora of other pairing mechanisms have been proposed and investigated \cite{HFsystems1984,1991SigristRMP,2017StewartAdvPhys}. One of the most influential extensions of the Cooper's idea comes from the observation that in the case of imbalanced systems, due to the mismatch of Fermi spheres of different components, the formation of correlated pairs forced by attractive mutual interactions is inseparably connected with a resulting non-zero net momentum of the pair \cite{fflo-ff1964,fflo-lo1964}. This unconventional pairing mechanism named after Fulde, Ferrell, Larkin, and Ovchinnikov (FFLO) has been extensively examined theoretically, mostly in the case of various solid-state systems such as iron-based superconductors \cite{FFLOIron1,FFLOIron2,FFLOIron3,FFLOIron4,kasahara2019evidence}, heavy-fermion compounds \cite{FFLOIron4,FFLOHeavy1,FFLOHeavy2,FFLOHeavy3,FFLOHeavy4}, or organic conductors \cite{FFLOOrganic1,FFLOOrganic2,FFLOOrganic3}.
However, it is also viewed as one of the possible ways to understand the fundamental properties of neutron stars \cite{alford2001colour,cirigliano2011low,2019SedrakianEPJA}, specific quantum chromodynamics models \cite{casalbuoni2005ginzburg}, or fermionic ultra-cold gases \cite{liao2010spin}. The latter example is of high importance since ultra-cold atomic systems, due to their tremendous tunability, are believed to be the best candidates for experimental observations of the FFLO state. Unfortunately, to date, the FFLO state is ephemeral and there are only indirect signs of this state of matter (see \cite{2018KinnunenRPP} for a recent review). Therefore, new theoretical approaches are proposed to capture the correlations, including the usage of bosons~\cite{singh2019enhanced}, and dynamical processes~\cite{kawamura2020nonequilibrium}.

\begin{figure}[t]
 \includegraphics[width=0.44\textwidth]{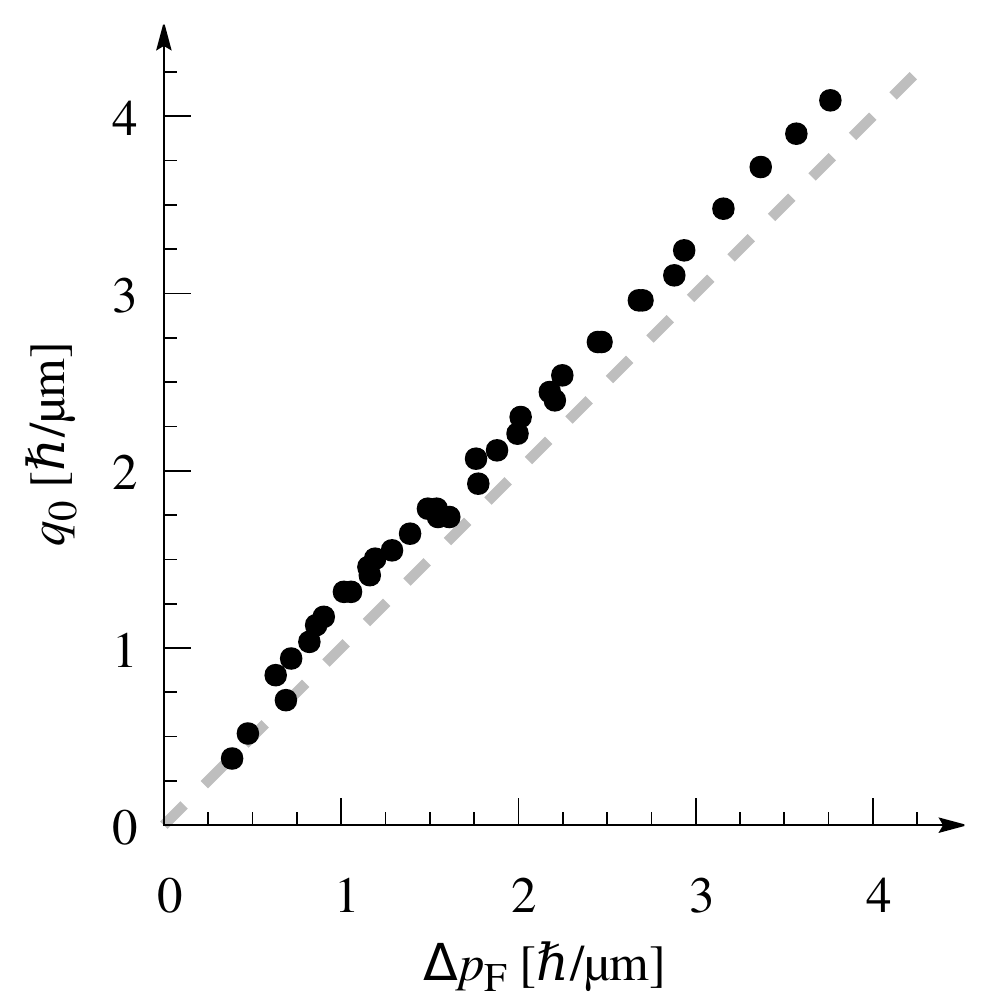}
 \caption{\label{fig:linear}
 The most probable FFLO momentum $q_0$ as a function of the Fermi momenta mismatch $\Delta p_\mathrm{F}$. Different points correspond to different number of particles and different imbalances. Exact description of each point is given in Fig.~\ref{FigA1} in the appendix. For clarity we do not show the point $q_0=\Delta p_F=0$ corresponding to the balanced scenario ($N_\uparrow=N_\downarrow$). The gray dashed straight line guiding the eye corresponds to phenomenologically predicted relation $q_0 = \Delta p_F$. Visible deviations from this prediction are ramifications of the finite number of particles and simplifications explained in the main text. }
\end{figure}

In this article, we show that the many-body ground-state of a few $^6$Li atoms confined in a harmonic trap (in the presence of mutual attractions) possesses many characteristic properties of the FFLO state which can be experimentally captured. For example, if one would combine recent progress in preparing spin-imbalanced few-fermion systems \cite{serwane2011deterministic} with the recently achieved development in measuring correlations between opposite spin fermions \cite{bergschneider2018correlations}, and perform the theoretical analysis of the obtained data along the recipe described here, then the most notable hallmark of the FFLO phase can be observed --- the {\it direct linear relation} between the most probable net momentum of the pair $q_0$ and the momentum mismatch between Fermi surfaces $\Delta p_F$ (see Fig.~\ref{fig:linear} with predictions for different numbers of particles and different spin-imbalance). Concurrently it should be emphasized that, in contrast to recent proposal~\cite{conduit2013fflo}, our approach is based on quantities which can be directly measured with nowadays techniques and does not require any significant modifications to the experimental setups. 

\section{Model}
Although our approach is very general and can be adopted to different fermionic systems confined in one-dimensional traps, we focus on a particular experimental realization --- the few-fermion mixture of $^6$Li atoms achieved currently almost on demand in Heidelberg \cite{serwane2011deterministic}. From a theoretical point of view, the system can be well described with the second-quantized Hamiltonian of the form
\begin{multline}\label{eq:ham}
 \hat{H} = \sum_{\sigma}
 \int\!\!\mathrm{d}{x}\,\hat{\Psi}^\dag_\sigma(x) \left( - \frac{\hbar^2}{2m} \frac{\mathrm{d}^2}{\mathrm{d}x^2} + \frac{1}{2}m\omega^2 x^2 \right) \hat{\Psi}_\sigma(x) \\
 + g \int\!\!\mathrm{d}{x}\,\hat{\Psi}_\uparrow^\dag(x) \hat{\Psi}_\downarrow^\dag(x) \hat{\Psi}_\downarrow(x)\hat{\Psi}_\uparrow(x),
\end{multline}
where $\omega\approx2\pi \cdot 1.488\,\mathrm{kHz}$ is the frequency of the external harmonic trap, $m$ is the mass of a $^6$Li atom, and $g$ is the effective one-dimensional interaction strength \cite{Olshanii1998}. The latter can be experimentally tuned by changing an external magnetic field and particularly it can become negative (effectively attractive interactions) \cite{zurn2013Pairing}. In the following we assume that $g$ is fixed by the external magnetic field $B=1202\,\mathrm{G}$ (see Table III in \cite{zurn2013Pairing}). If one expresses all quantities in natural units of the harmonic oscillator, {\it i.e.}, energies in $\hbar\omega=9.86\,\cdot 10^{-31}\,\mathrm{J} $, positions in $\sqrt{\hbar/m\omega}=1.06\,\mathrm{\mu m}$, and wave vectors in $\sqrt{m \omega/\hbar}=0.95\,\mathrm{\mu m}^{-1}$ then the assumed interaction strength corresponds to $g=-1$. The fermionic field operator $\hat{\Psi}_\sigma(x)$ annihilates a $\sigma$-component fermion at a position $x$ and obeys standard anti-commutation relations $\{\hat{\Psi}_\sigma(x),\hat{\Psi}_{\sigma'}^\dag(x')\}=\delta(x-x')\delta_{\sigma\sigma'}$. For further convenience, we introduce density operators in the position and momentum representations, $\hat{\rho}_\sigma(x) = \hat\Psi_\sigma^\dagger(x)\hat\Psi_\sigma(x)$ and $\hat{\pi}_\sigma(p) = \hat\Psi_\sigma^\dagger(p)\hat\Psi_\sigma(p)$, where $\hat\Psi_\sigma(p)=\int\!\mathrm{d}x\,\hat{\Psi}_\sigma(x)\mathrm{exp}(-i p x/\hbar)$ is a Fourier transform of the field operator $\hat\Psi_\sigma(x)$.

To perform appropriate calculations for a given number of particles $N_\uparrow$ and $N_\downarrow$, we express the Hamiltonian \eqref{eq:ham} as a matrix in the Fock basis of many-body eigenstates of the non-interacting system $\{|F_i\rangle\}$. The basis is given as a set of products of different Slater determinants of $N_\uparrow$ and $N_\downarrow$ harmonic potential orbitals chosen appropriately for each component. Since the Hilbert space grows exponentially along with the number of particles and number of single-particle orbitals, we restrict ourselves only to these Fock states which have the non-interacting energy lower than some properly chosen cut-off. As shown recently, as long as we are interested in the many-body ground state of the system, this approach can be applied effectively for any trapping potential and any number of particles \cite{1998HaugsetPRA,2007DeuretzbacherPRA,2018PlodzienARX,Chrostowski2019,Pecak2019Intercomponent}. The resulting matrix representation of the many-body Hamiltonian \eqref{eq:ham} is diagonalized using the Arnoldi method~\cite{ARPACK1998Sorensen} and the many-body ground-state $|G_0\rangle$ is found as its decomposition coefficients in the non-interacting basis $\{|F_i\rangle\}$. 

\section{Results}
Pairing between opposite component fermions, even if actually present in the system, is very resistant to detection. In principle, it requires experimental access to all possible two-particle measurements, {\it i.e.}, a complete two-particle reduced density matrix (2RDM) is needed. Recently there were many attempts to measure inter-particle correlations in different scenarios~\cite{sherson2010single,Cheuk2015FermionicMicroscope,ParsonsPRL2015,Cheuk2016Correlations,OmranBloch2015PauliBlocking,Andrea2018Imaging} but all of them give access only to the diagonal parts (two-particle density profiles) rather than a complete 2RDM. 
Therefore some other theoretical framework is needed to capture the mutual correlations. Fortunately, it was argued \cite{lukin2004noisCorr,Altman2009lowDim,Altman20081D,folling2005spatial} that higher-order correlations induced by interactions can be extracted from pure diagonal parts of 2RDM by an appropriate subtraction of spurious correlations. Spurious correlations arise also in the absence of interactions, independently of the quantum statistics, and depend only on single-particle densities. 
In the case of a two-component mixture of distinguishable fermions, the so-called two-point noise correlation $\cal G$ is a convenient tool to unravel the quantum correlations from the trivial background \cite{lukin2004noisCorr,2008LuscherPRA,brandt2017two, Pecak2019Intercomponent}. It is defined respectively in the position and momentum representation straightforwardly as:
\begin{subequations} \label{eq:noise}
\begin{align} 
 {\cal G}_\rho(x_\uparrow;x_\downarrow) &= \langle \hat{\rho}_\uparrow(x_\uparrow)\hat{\rho}_\downarrow(x_\downarrow)\rangle-\langle \hat{\rho}_\uparrow(x_\uparrow)\rangle\langle \hat{\rho}_\downarrow(x_\downarrow)\rangle,\label{eq:NoiseX} \\
 {\cal G}_\pi(p_\uparrow;p_\downarrow) &= \langle \hat{\pi}_\uparrow(p_\uparrow)\hat{\pi}_\downarrow(p_\downarrow)\rangle-\langle \hat{\pi}_\uparrow(p_\uparrow)\rangle\langle \hat{\pi}_\downarrow(p_\downarrow)\rangle.\label{eq:NoiseP}
\end{align}
\end{subequations}
Note that in the non-interacting limit ($g\rightarrow 0$) the noise correlations \eqref{eq:noise} identically vanish. Therefore they can be interpreted as quantities measuring the amount of two-body correlation in the system forced purely by interactions. Importantly, it should be emphasized at this point that exactly this kind of correlations was captured experimentally very recently \cite{bergschneider2018correlations} in an optical lattice and can be applied to continuum systems as well.
\begin{figure}
 \includegraphics[width=0.49\textwidth]{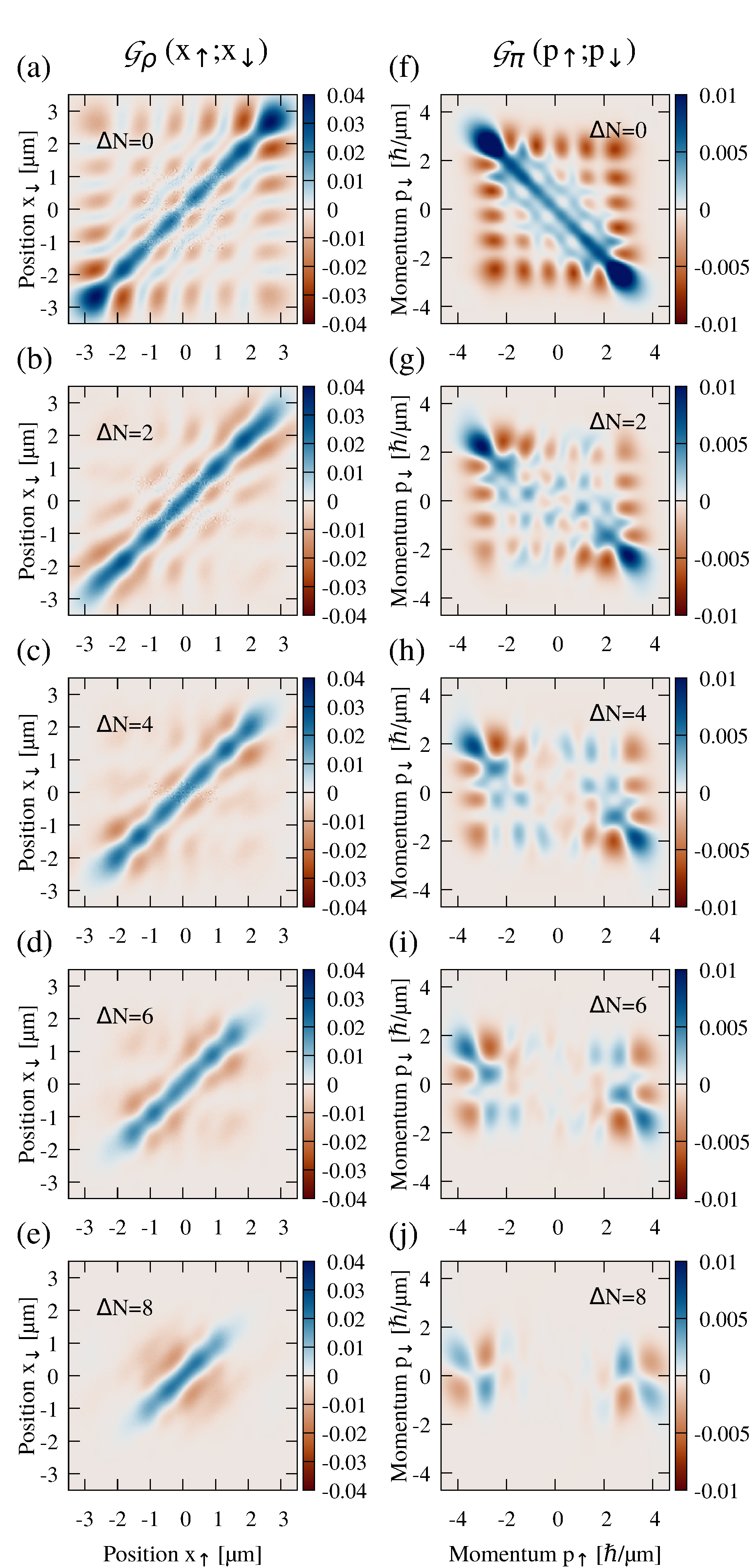}
 \caption{\label{fig:noise}
 The noise correlation $\cal{G}_\rho$ ($\cal{G}_\pi$) in the position (momentum) domain is presented in the left (right) column. (a--e) Attractive interactions enhance the probability of finding particles of two species in the same position for different configurations of $N=N_\uparrow+N_\downarrow=10$ particles. (f) Simultaneously, the momenta are anti-correlated for the same number in both components $N_\uparrow=N_\downarrow$. (g--j) Whenever the particle imbalance $\Delta N\neq 0$ is introduced, there is a visible shift in momentum that corresponds to the net momentum of a correlated pair.}
\end{figure}

In Fig.~\ref{fig:noise} we plot noise correlations \eqref{eq:noise} for the system with $N_\uparrow+N_\downarrow=10$ particles and different imbalances $\Delta N=N_\uparrow-N_\downarrow$. Without losing generality, in the following, we consider only non-negative imbalances $\Delta N\geq0$. In the balanced case $\Delta N=0$ (Figs.~\ref{fig:noise}(a) and \ref{fig:noise}(f)) the Fermi spheres for both components are exactly the same and the standard Cooper-pairing mechanism occurs in the system~\cite{Sowinski2015Pairing}. Consequently, when the noise correlation is considered, the pairing mechanism is manifested by a strong anti-correlation of the fermions' momenta --- strong enhancement of the probability of finding fermions with exactly opposite momenta $p_\uparrow=-p_\downarrow$ is clearly evident. This picture is substantially changed when the particle imbalance is introduced to the system (see Figs.~\ref{fig:noise}(g--j)). It is quite evident that in these cases the anti-diagonal enhancement of correlations is split into two ridges which are pushed out from the line $p_\uparrow+p_\downarrow=0$. It means that in contrast to the balanced scenario the most probable outcome of the two-point momentum measurement is that paired fermions have nonzero net momentum $q_0=p_\uparrow+p_\downarrow\neq 0$. It is also very clear that the total momentum $q_0$ monotonically increases with the imbalance $\Delta N$ which is one of the clearest signatures of FFLO-like pairing. 
Moreover, in the balanced case ($\Delta N=0$) all momenta (below maximal Fermi momentum) are accessible for fermions and they almost equally contribute to the collective pairing mechanism (notice an almost flat distribution along the anti-diagonal for $\Delta N=0$ in Fig.~\ref{fig:noise}(f)). However, when the system is imbalanced, particles with smaller momenta do not contribute to the formation of pairs with net momentum $q_0$ (the empty region in the middle of the noise correlation ${\cal G}_\pi$ for $\Delta N>0$ in Figs.~\ref{fig:noise}(g--j)). This kind of formation of correlated pairs is another well-known property of the FFLO mechanism \cite{2008LuscherPRA,2018KinnunenRPP}.

\begin{figure}
\includegraphics[trim={0 0 0 0},clip,width=0.45\textwidth]{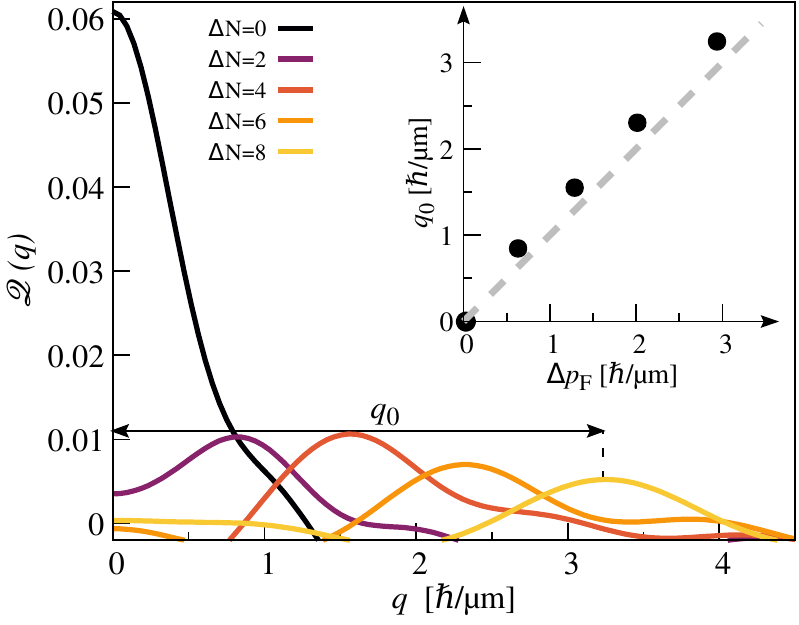}
 \caption{\label{fig:Qq}
 Distribution of the net FFLO momentum $\mathcal{Q}(q)$ for the system of $N=10$ particles and subsequent particle imbalances $\Delta N=0,2,4,6,8$. Clear maxima in momenta correspond to the most probable net momentum $q_0$ of the correlated pair for a given imbalance. (Inset) The most probable net momentum $q_0$ as a function of the Fermi momenta mismatch $\Delta p_{\mathrm{F}}$. Results for different particle numbers $N$ are aggregated in Fig.~\ref{fig:linear}.}
\end{figure}

In the next step we aim to find the most probable net momentum of the pair $q_0$ as a function of the imbalance $\Delta N$. For this purpose, we introduce the filtering procedure giving us the possibility to quantify the occurrence of different FFLO momenta $q$. In general, the filtering is done by convoluting the noise correlation with an appropriately chosen filter function:
\begin{equation}
{\cal Q}(q) = \int \mathrm{d}p_\uparrow\mathrm{d}p_\downarrow\,{\cal F}(p_\uparrow+p_\downarrow+q) \mathcal{G}_\pi(p_\uparrow;p_\downarrow).
\end{equation}
In our approach we choose the simplest Gaussian filtering function ${\cal F}(\xi) = (\pi\kappa)^{-1/2}\mathrm{exp}(-\xi^2/2\kappa^2)$ with $\kappa$ being of the order of the perpendicular width of the enhanced correlations area. We checked that the final results are not sensitive to the exact shape of the filtering function, since for reasonable values of $\kappa$ the most probable momentum $q_0$ (the value for which the measure ${\cal Q}(q)$ is maximum) does not change. In Fig.~\ref{fig:Qq} we plot the resulting function ${\cal Q}(q)$ for the system of $N=10$ particles and different imbalances $\Delta N$. It is clear, that the balanced system ($\Delta N=0$) is characterized by vanishing $q_0$ (black curve). When the particle imbalance is increased, the maximum moves towards higher absolute values of the momenta. 

Finally, to make a whole picture comprehensive, we make a connection of the imbalance $\Delta N$ with the discrepancy between Fermi momenta of both components $\Delta p_F$. In the case of an essentially non-homogenous system of a few particles, the definition of the Fermi momentum is obviously not straightforward since the system is not translationally invariant. Particularly, it is no longer valid that the Fermi momentum $p_{F\sigma}$ is proportional to the number of particles $N_\sigma$. Moreover, due to a small number of particles, any approaches based on the local density approximation being appropriate for a large number of particles (see for example \cite{2007OrsoPRL,2008LuscherPRA,2009EdgePRL}) are also not adequate. To overcome this difficulty, let us first notice that the well-determined quantity is the Fermi energy in the limit of weak interactions, $\epsilon_{F\sigma}=\hbar\omega(N_\sigma-1/2)$. This energy defines the maximal value of the momentum which is accessible for the particle moving on the Fermi surface, $p_{F\sigma}=\sqrt{2m\epsilon_{F\sigma}}$. In the semi-classical picture, this is a momentum gained by a particle when it passes through the middle of the trap. If we associate the Fermi momentum with this quantity then we immediately find a phenomenological connection between the imbalance $\Delta N$ and the maximal discrepancy of the Fermi momenta $\Delta p_F$. When momenta are expressed in the natural unit $\sqrt{\hbar m\omega}$ then this relation reads
\begin{equation}\label{eq:mismatch}
\Delta p_\mathrm{F} \approx p_{F\uparrow} - p_{F\downarrow} = \sqrt{2\left(N_\uparrow - \frac{1}{2}\right)} - \sqrt{2\left(N_\downarrow - \frac{1}{2}\right)}.
\end{equation}
Applying this definition, in the inset of Fig.~\ref{fig:Qq} we plot the most probable net momentum of the pair $q_0$ as a function of the discrepancy $\Delta p_{\mathrm{F}}$ for $N=10$. It is clearly evident that all points lie almost exactly on the straight line. The situation is very similar if one repeats this procedure for a different number of particles. In Fig.~\ref{fig:linear} we show numerical results for $N=3,\ldots,14$ and different imbalances $\Delta N$. All these points almost ideally support the relation $q_0\approx \Delta p_F$ (dashed line) and display one of the fundamental consequences of the FFLO pairing mechanism --- the net momentum of the Cooper pair is equal to the Fermi momenta discrepancy, $q_0 = \Delta p_F$. 

Finally, let us discuss the evident deviations between numerical results and predictions of our phenomenological derivation. They can be explained on three levels. {\it (i)} It is clear that the definition of the momentum mismatch $\Delta p_F$ is very phenomenological and simplified. It focuses only on one momentum associated with the Fermi level. Therefore it may predict only a general trend rather than an exact relation. {\it (ii)} It is known that the relation between the FFLO momentum and the components' Fermi momenta is, in fact, more complicated when interactions and an effective pairing potential are taken into account. For example, as discussed in \cite{2007KaczmarczykAPPA}, even a simplified inclusion of an effective pairing potential immediately leads to increasing of the FFLO momentum. This effect is clearly seen in Fig.~\ref{fig:linear} (all points are shifted towards larger $q_0$). Moreover, as clearly evident in Fig.~\ref{fig:Qq}, the distribution of possible FFLO momenta becomes very wide for larger imbalances. Therefore, choosing a single $q_0$ to characterize a whole distribution is evidently oversimplified. {\it (iii)} The theory of FFLO pairing explains the appearance of correlations in terms of the collective cooperation of all particles in the system. Therefore, it gives rigorous relations only for a large number of particles. From this point of view, the existence of some finite-size corrections is quite natural. They lead to small shifts of particular points in Fig.~\ref{fig:linear}. 

\section{Conclusions}
In summary, we showed --- based on exact numerical calculations --- that in the confined one-dimensional system of a few attractively interacting fermionic $^6$Li atoms the FFLO pairing mechanism is clearly manifested and {\it can be detected} with current experimental techniques. Taking into account the tremendous tunability of ultra-cold systems, our proposal not only opens another route for the {\it direct} experimental confirmation of unconventional pairing forced by broken symmetry between components, but also reveals an additional tool for studying the appearance of collectiveness when the quantum system undergoes a transition from the few to the many-body limit. At this point, we want also to mention that the FFLO mechanism can be considered for spin-balanced few-body systems but with different mass atoms \cite{2019Lydzba}. Taking into account the huge experimental progress in controlling mass-imbalanced Fermi mixtures \cite{cetina2016ultrafast,Grimm2018DyK,neri2020CrLi}, this possibility is also in-game. It should be noted, however, that the few-body regime for these kinds of systems has not yet been achieved.  

\section*{Acknowledgements}
The authors are very grateful to Konrad Kapcia, Maciej Lewenstein, Patrycja {\L}yd\.zba, and Piotr Magierski for their fruitful comments and inspiring questions at different stages of this project. The authors are indebted to the FoKA community for many inspiring discussions. This work was supported by the (Polish) National Science Center Grants No. 2017/27/B/ST2/02792 (DP) and 2016/22/E/ST2/00555 (TS). Numerical calculations were partially carried out in the Interdisciplinary Centre for Mathematical and Computational Modeling, University of Warsaw (ICM) under the computational grant No. G75-6.

\appendix*
\section{}
\begin{figure}[h!]
\centering
\includegraphics[width=0.9\linewidth]{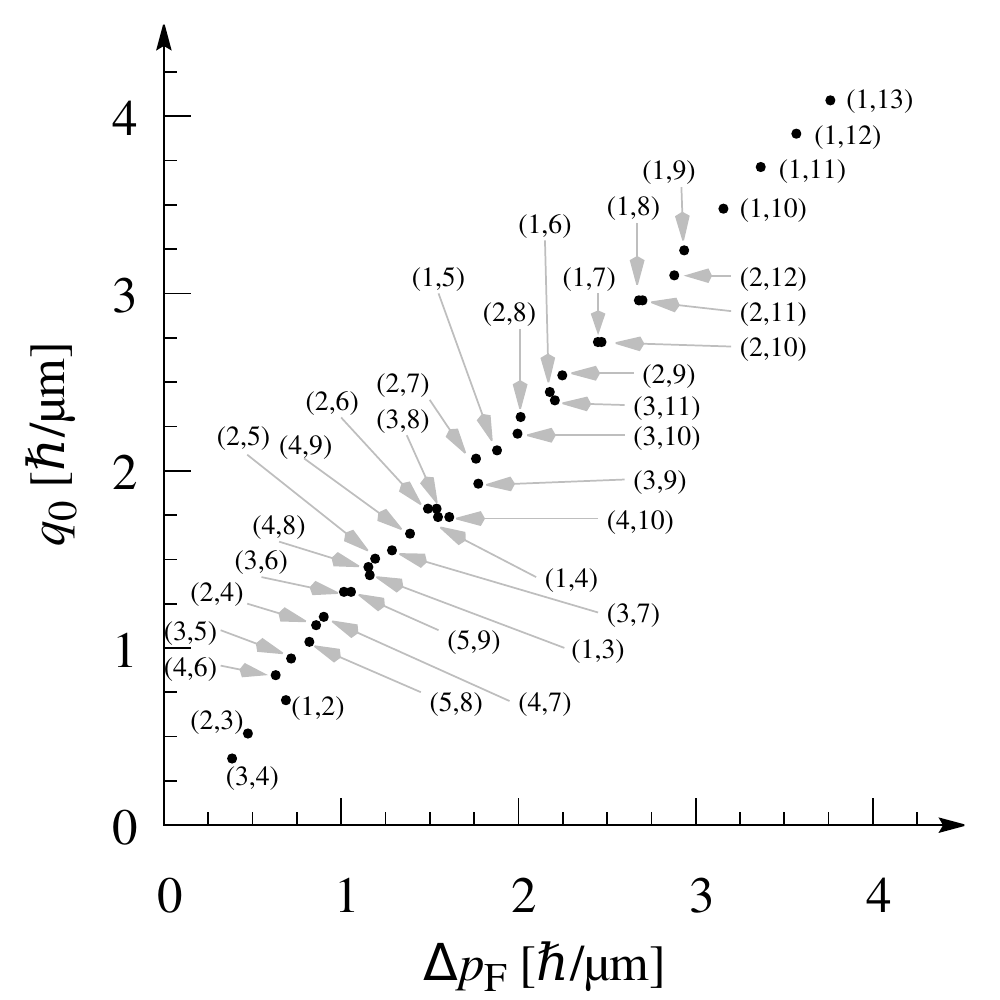}
\caption{  
 The most probable FFLO momentum $q_0$ as a function of the Fermi momenta mismatch $\Delta p_\mathrm{F}$. Different points correspond to a different number of particles and different imbalances. Labels linked to particular points denote the number of particles $(N_\downarrow,N_\uparrow)$. For clarity we do not show the point $q_0=\Delta p_F=0$ corresponding to the balanced scenario ($N_\uparrow=N_\downarrow$). 
\label{FigA1} }
\end{figure} 

As explained in the main text, Fig.~\ref{fig:linear} shows a linear relation between the most probable net momentum of the pair $q_0$ and the momentum mismatch of Fermi spheres $\Delta p_F$. The universality of this relation is supported by exact numerical results obtained for different configurations $(N_\uparrow,N_\downarrow)$ of the system studied. Each configuration is represented by a black dot on the plot. In Fig.~\ref{FigA1} we redraw these results and explicitely give labels denoting these configurations.
\bibliographystyle{apsrev4-1}
\bibliography{FEWbibtex}

\end{document}